\documentclass[10pt,aps,prc,superscriptaddress,twoside,twocolumn,nofootinbib,showpacs]{revtex4-2}
\usepackage{amsmath,amssymb}
\usepackage{mathtools}
\usepackage{bbold}
\usepackage[dvipsnames]{xcolor}
\usepackage{braket}
\usepackage{graphicx}
\usepackage{slashed}
\usepackage{booktabs}
\usepackage{tensor}
\usepackage{multirow}
\usepackage{mwe}
\usepackage{changes}
\usepackage{enumerate}
\usepackage{stackrel}
\usepackage{physics}
\usepackage{bm}
\usepackage{multirow}

\allowdisplaybreaks

\newcommand{\np}{$np$ }

\usepackage{epsfig}
\usepackage{subfigure}

\allowdisplaybreaks

\begin{document}
\title{Data-Driven Statistical Ensembles of Chiral Nuclear Interactions}

\author{Pengsheng \surname{Wen}}
\email{pswen2019@physics.tamu.edu}
\affiliation{Cyclotron Institute, Texas A\&M University, College Station, TX 77843, USA}
\affiliation{Department of Physics and Astronomy, Texas A\&M University, College Station, TX 77843, USA}

\author{Jeremy W. \surname{Holt} }
\email{holt@physics.tamu.edu}
\affiliation{Cyclotron Institute, Texas A\&M University, College Station, TX 77843, USA}
\affiliation{Department of Physics and Astronomy, Texas A\&M University, College Station, TX 77843, USA}

\begin{abstract}
Recent advances in ab initio nuclear theory, machine learning, and Bayesian inference, coupled with increasingly precise nuclear experiments and astrophysical observations, have enabled more robust constraints on fundamental descriptions of the nuclear interaction.
Although well-established nonlinear regression methods can identify best-fit sets of low-energy constants at fixed resolution scale, they provide limited insight into the full underlying probability distributions of those constants. 
A central remaining challenge in nuclear theory is therefore to characterize full probability distributions of nuclear forces across resolution scales.
In this work, we employ normalizing flows, a class of expressive generative machine learning models, to infer the joint probability distribution of two-body low-energy constants (LECs) in chiral effective field theory over a wide range of resolution scales. 
The resulting LEC distributions are shown to accurately reproduce experimental neutron–proton scattering phase-shift distributions.
Furthermore, strong non-Gaussian correlations among LECs are revealed, indicating a nontrivial interplay among distinct short-range nuclear dynamics.
This work establishes a general framework for constructing statistical ensembles of nuclear interactions that can be systematically constrained by future nuclear experiments and astrophysical observations.

\end{abstract}

\maketitle

{\it Introduction:}
Developing accurate microscopic models of the nuclear force has been a central challenge in nuclear physics for more than half a century. Nuclear potentials are not themselves observables \cite{aokiLatticeQCDBaryonBaryon2020,ishiiNuclearForceLattice2007} and are therefore not uniquely determined by experimental data, which introduces systematic uncertainties associated with their parameterization and resolution scale \cite{dyhdaloRegulatorArtifactsUniform2016,Wen:2023oju}. Chiral effective field theory ($\chi$EFT) \cite{machleidtChiralEffectiveField2011,epelbaumModernTheoryNuclear2009} has provided a systematic framework for describing the nuclear force in terms of long-range pion-exchange processes and short-range parameterized contact interactions. The short-distance interactions are governed by low-energy constants (LECs) that must be determined from experimental data and depend on the resolution scale of the interaction. A comprehensive characterization of the nuclear force therefore requires not only optimal values of these parameters, but their correlated probability distributions across resolution scales.

Traditionally, LECs have been determined at fixed resolution scales through nonlinear regression \cite{Wild:2014swa,10.1007/BFb0067700}, 
typically by fitting nucleon-nucleon scattering data and few-body observables \cite{PhysRevC.68.041001,PhysRevLett.122.042501,PhysRevC.91.051301,PhysRevLett.115.122301,PhysRevLett.110.192502,PhysRevX.6.011019,Hu:2025cjl}. More recently, Bayesian inference and Markov chain Monte Carlo methods have enabled correlated multivariate LEC distributions incorporating experimental and theoretical uncertainties \cite{PhysRevC.107.014001,PhysRevC.105.014004,carlssonUncertaintyAnalysisOrderbyOrder2016}, while complementary approaches have constructed families of nuclear interactions continuously across resolution scales \cite{Wen:2023oju}. These developments address two distinct sources of uncertainty in the nuclear interaction, but a unified characterization of correlated LEC distributions over a continuous range of resolution scales remains lacking. Such a framework becomes especially important as increasingly diverse constraints from nucleon-nucleon scattering, finite nuclei \cite{ekstromOptimizedChiralNucleonNucleon2013,epelbaumPrecisionNucleonNucleonPotential2015,PhysRevLett.126.172502,Hu:2021trw} and neutron-star observations \cite{9hjh-k5wm,galaxies10050099} are incorporated into fits of LECs. In addition, experimental observables are not always represented by single values, but may follow distributions that encode intrinsic physical fluctuations and measurement uncertainties. 

In the present work we develop a generative machine-learning framework to infer complete joint probability distributions of two-body $\chi$EFT LECs continuously across resolution scales. We employ normalizing flows (nflows) \cite{gaoFlowHighdimensionalIntegration2020,papamakariosNormalizingFlowsProbabilistic2021,bradyNormalizingFlowsMicroscopic2021,wenApplicationNormalizingFlows2024} whose expressive transformations enable efficient modeling and sampling of complex, high-dimensional probability distributions. As a proof of concept, we constrain these distributions using neutron-proton scattering phase shifts and demonstrate that the resulting ensembles of nuclear interactions reproduce the corresponding experimental error distributions. The inferred LEC distributions exhibit strong correlations that evolve continuously with resolution scale. More broadly, the framework provides a means of synthesizing nuclear experimental and astrophysical observational data to construct statistically constrained ensembles of nuclear interactions.

{\it Normalizing flow and fitting framework:}
Normalizing flows are a class of generative machine learning models that employ expressive and trainable transformations to map high-dimensional samples from a simple latent distribution to a complex target distribution. This capability makes them well-suited for modeling intricate probability distributions and efficiently generating samples that follow the desired distribution. 
In this work, we employ normalizing flows to model the probability distributions of the LECs in the next-to-next-to-next-to-leading order (N3LO) chiral nuclear potentials of Ref.~\cite{PhysRevC.91.014002}. 

A normalizing flow model uses a chain of transformation layers to transform samples from the latent space to the target space. 
The transformation can be formally expressed as 
\begin{align}
    \bm{x}_L = f_{L-1} \circ f_{L-2} \circ \dots \circ f_i \circ \dots\circ f_0(\bm{x}_0),
\end{align}
where $\bm{x}_L$ is the sample in target space, $\bm{x}_0$ is the sample in latent space, and $f_i$ is the transformation function in the $i$th layer.
Transformation functions are designed to be analytic, so that the distribution of $\bm{x}$ can be calculated using 
\begin{align}
    p(\bm{x}_L)
    &= q(\bm{x}_0) \prod_i \left\| \frac{\partial f_i}{\partial \bm{x}_{i}} \right\|^{-1},
\end{align}
where $\bm{x}_i = f_i(\bm{x}_{i-1})$ is the output at $i$th layer, and $\| {\partial f_i}/{\partial \bm{x}_{i}}\| $ denotes the absolute value of the determinant of the Jacobian of this transformation.
The initial distribution $q(\bm{x}_0)$ is chosen to be simple so that the latent space sample $\bm{x}_0$ can easily be generated. 
In our work, we adopt a uniform distribution as $q(\bm{x}_0)$. 
The transformation functions $f_i$ are constructed to be expressive and parameterized by trainable variables, enabling the final distribution $p(\bm{x}_L)$ to approximate a complicated target distribution.
The trainable parameters in the $f_i$ can be designed to be dependent on an external condition $\bm{z}$, yielding a conditional distribution $p({\bm x}_L|\bm{z})$.
We employ rational quadratic splines as our transformation function, which splits the sampling space into multiple bins parameterized by trainable bin widths, bin heights, and knot derivatives.
All the trainable parameters are allowed to depend on the external condition ${\bm z}$. 
In this work, the resulting samples $\bm{x}_L$ are vectors of $\chi$EFT LECs, with the resolution scale $\Lambda$ serving as the conditioning variable ${\bm z}$.

For an experimental target, we focus on the neutron-proton ($np$) scattering phase shifts and their experimentally derived uncertainties as observables to infer the distribution of LECs. 
Propagating a parameter distribution to observables is particularly challenging when the forward model is nonlinear and computationally expensive, since the resulting observable distribution generally has no closed-form expression.
We therefore determine the observable distribution numerically: LEC samples generated by the normalizing flow are propagated to the corresponding phase shifts, whose probability density is estimated using a Gaussian kernel density estimator (KDE) \cite{Davis2011,Parzen1962OnEO}.

The discrepancy between the estimated probability distribution function (PDF) and the target experimental distribution of an observable, serving as the loss function for optimizing the normalizing flow, is quantified by the reverse Kullback-Leibler (KL) divergence \cite{csiszarIDivergenceGeometryProbability1975,bishop2006prml}, which is a widely used objective in machine learning for inferring probability distributions and fitting generative models:
\begin{align}\label{eq:loss}
    {\rm loss} &= 
    \sum_{\bm{z}}
    \sum_{\mathcal{O}}
    \frac{1}{N}\sum_{i=1}^N
    \log \left(
        \frac{p_{\rm theor}(\mathcal{O}_i|\bm{z})}{p_{\rm exp}(\mathcal{O}_i)}
    \right),
\end{align}
where $i$ is the index for $i$-th sample (i.e., a specific set of LECs), $\mathcal{O}_i$ denotes the calculated observable corresponding to the sample, $p_{\rm exp}$ is the target PDF from experiment, and $p_{\rm theor}$ represents the estimated PDF obtained from the KDE, which can be a conditional distribution of ${\bm z}$ if the distribution of ${\bm x}_L$ is dependent on ${\bm z}$. 
The ${\rm loss}$ is summed over all the types of observables, such as different phase shifts and nuclei properties, and external conditions considered in the analysis. 

In this work, the LECs are grouped according to their corresponding phase shift channels, and each channel is fitted independently. 
For a given channel, the observables are taken to be the \np phase shifts at laboratory energies $E_{\rm lab}$ up to $200$ MeV.
We take the mean values and standard deviations from the partial wave analysis of Ref.~\cite{NavarroPerez:2014ihw} as the experimental reference.
The sum over $\mathcal{O}$ corresponds to the sum over all selected values of $E_{\rm lab}$. 
The normalizing flow is trained by minimizing a loss function that encourages the theoretical distribution to approximate the experimental one. 
Once the distributions of the calculated and experimental observables are similar, the normalizing flow successfully finds a suitable LEC distribution.
Compared with traditional non-linear regression algorithms, in which the optimization is driven by the deviation of predicted values corresponding to a single sample, our approach defines the optimization direction through the discrepancy between two distributions $p_{\rm theor}$ and $p_{\rm exp}$.
This requires a sufficiently large number of samples to ensure that the estimated distribution from KDE is accurate, which is well-suited for normalizing flows, as they can generate large batches of samples simultaneously.
By focusing on distribution discrepancies, the optimization captures global features of the parameter space, rather than searching for a single optimal solution in conventional nonlinear regression fitting.
Moreover, distributions across all relevant conditions, which are the $\Lambda$s in $\chi$EFT, are summed and optimized simultaneously, making the framework well suited for comprehensive uncertainty quantification of $\chi$EFT.

The calculation of phase shifts {$\delta$} from LECs at a given $\Lambda$ is nonlinear and computationally intensive. 
The procedure requires first constructing the nuclear potential for a given set of LECs and $\Lambda$, then performing a partial-wave decomposition for the potential, and finally solving the Lippmann–Schwinger (LS) equation to obtain the phase shifts.
To facilitate the fitting process, we trained another neural network ${\rm NN}^{(L,S,J)}_{\delta}$ as a phase shift simulator whose inputs are the LECs and $\Lambda$. 
Its outputs are the corresponding phase shift values at the chosen $E_{\rm lab}$ points and angular momentum channels $(L,S,J)$:
$\delta_{^{2S+1}L_J}^{E_{\rm lab}}({\rm LECs},\Lambda)$.

\begin{figure}[t]
    \centering
    \includegraphics[width=\linewidth]{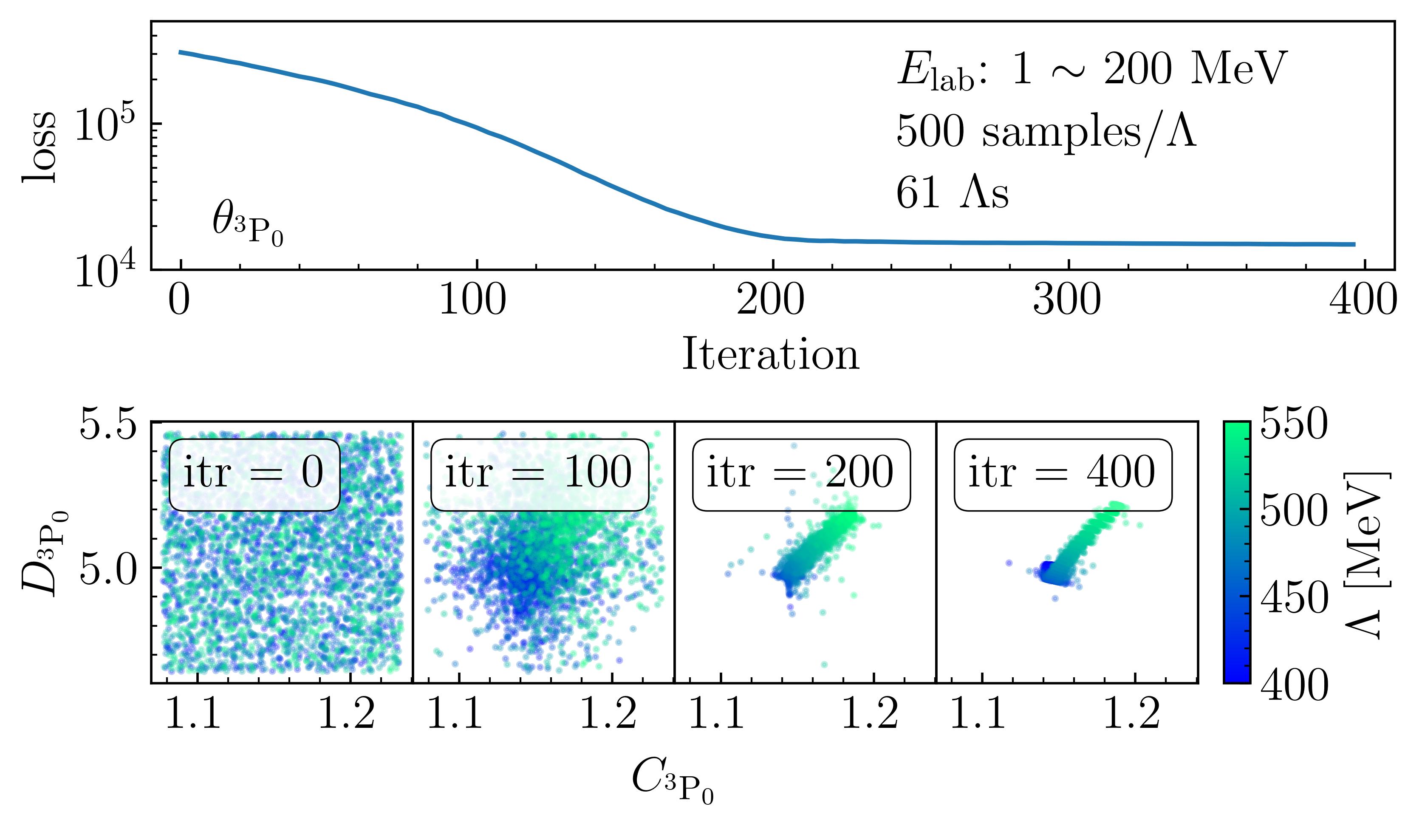}
    \caption{
    (Top panel) Normalizing flow training for the LECs ${C}_{^3{\rm P}_0}$ and ${D}_{^3{\rm P}_0}$ at N3LO in $\chi$EFT with 61 values of $\Lambda$ ranging from 400 to 550 MeV.
    The loss function (top panel) is determined by comparing to phase shifts in the $^3{\rm P}_0$ channel and summed over the laboratory energies up to $200$ MeV. 
    (Bottom panel) Randomly generated samples from the normalizing flow at 0th, 100th, 200th, and 400th iteration. 
    }
    \label{fig:train}
\end{figure}
{\it Result and Discussion:}
We show the training of a normalizing flow that is used to model the distribution of the LECs $(C_{^{3}{\rm P}_0}, D_{^{3}{\rm P}_0})$ in Fig.~\ref{fig:train}.
In each training iteration, a batch of LEC samples with different values of the cutoff $\Lambda$ is generated from the normalizing flow. 
Their corresponding observables, the $^3{\rm P}_{0}$ \np phase shifts at the selected laboratory energies $E_{\rm lab}$, are calculated using a well-trained simulator. 
At a fixed $E_{\rm lab}$, the probability density of each simulated phase shift is estimated using Gaussian KDE, and the average logarithmic term in Eq.\ \eqref{eq:loss} over all samples represents the contribution from this energy to the overall loss function.
The target distribution is a Gaussian with the experimental mean and standard deviation at the corresponding $E_{\rm lab}$.
Summing the contributions over all selected energies and resolution scales $\Lambda$ yields the final value of loss function.

The loss function values during optimization are shown in the top panel of Fig.~\ref{fig:train}. 
The optimization proceeds successfully, as evidenced by the decrease of the loss function with training iterations, indicating that the phase-shift distributions produced by the LECs generated by the normalizing flow approach the target experimental distributions.
The bottom panel shows the generated samples of the LECs.
At the initial (0th) iteration, the samples are uniformly distributed for all the $\Lambda$s.
With the training iterations, samples collapse into a strongly correlated distribution, forming distinct clusters associated with different $\Lambda$ values.
The concentration of the distribution demonstrates that the normalizing flow successfully filters out non-physical regions of the parameter space for a given $\Lambda$. 
In addition, the correlations observed between specific pairs of LECs reveal the intrinsic interplay between the corresponding short-range interaction mechanisms, reflecting the possible underlying physics that gives rise to these terms.

\begin{figure}[t]
    \centering
    \includegraphics[width=\linewidth]{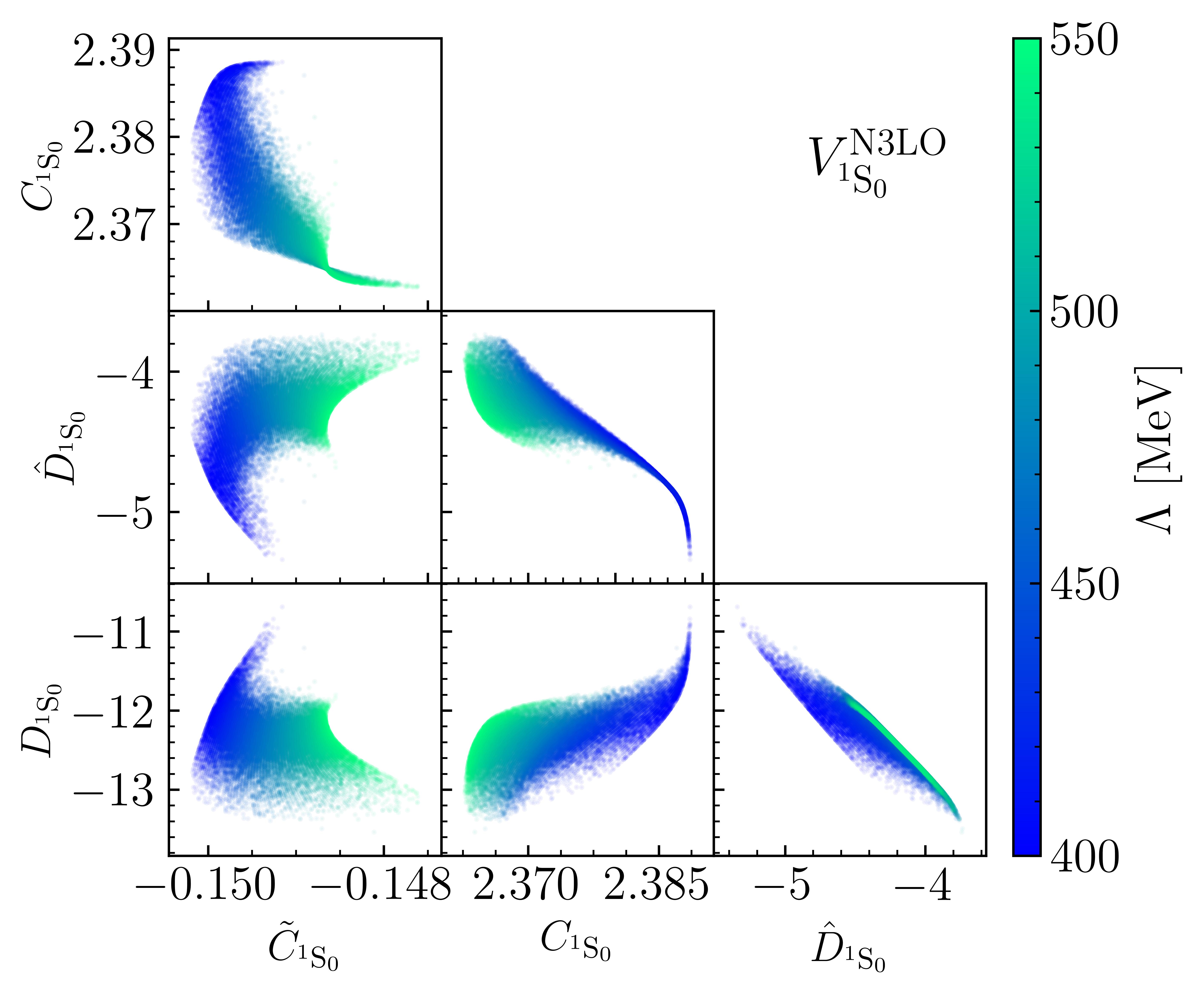}
    \caption{
    Joint distribution of the four $^1{\rm S}_0$ LECs $\tilde{C}_{^1{\rm S}_0}$, $C_{^1{\rm S}_0}$, $D_{^1{\rm S}_0}$, and $\hat{D}_{^1{\rm S}_0}$ at N3LO in $\chi$EFT from a trained normalizing flow model with cutoff variations in the range $400\,{\rm MeV} \le \Lambda \le 550\,{\rm MeV}$.
    }
    \label{fig:rholec}
\end{figure}
The optimized distribution of the LECs at N3LO for the $^1{\rm S}_0$ phase shift channel is shown in Fig.~\ref{fig:rholec}. 
At N3LO, the LECs in this phase shift channel include four components: $\tilde{C}_{^1{\rm S}_0}$, $C_{^1{\rm S}_0}$, $D_{^1{\rm S}_0}$ and $\hat{D}_{^1{\rm S}_0}$.
This four-dimensional parameter space is naturally more challenging to obtain an accurate distribution model than in the case of the $^3{\rm P}_{0}$ channel.
As shown in Fig.~\ref{fig:rholec}, the normalizing flow finds a concentrated and strongly correlated distribution among the LECs for each $\Lambda$. 
Even though the parameter space is high-dimensional, the normalizing flow successfully identifies the narrow, physically meaningful region of the LECs for each $\Lambda$ across a broad range of resolution scales (from 400 to 550 MeV).
The high precision of the experimental data tightly constrains the allowed parameter space, leading to the narrow distribution for each $\Lambda$. 
The normalizing flow reveals strong and complicated correlations among the four LECs, whose structure evolves with $\Lambda$.

\begin{figure}[t]
    \centering
    \includegraphics[width=\linewidth]{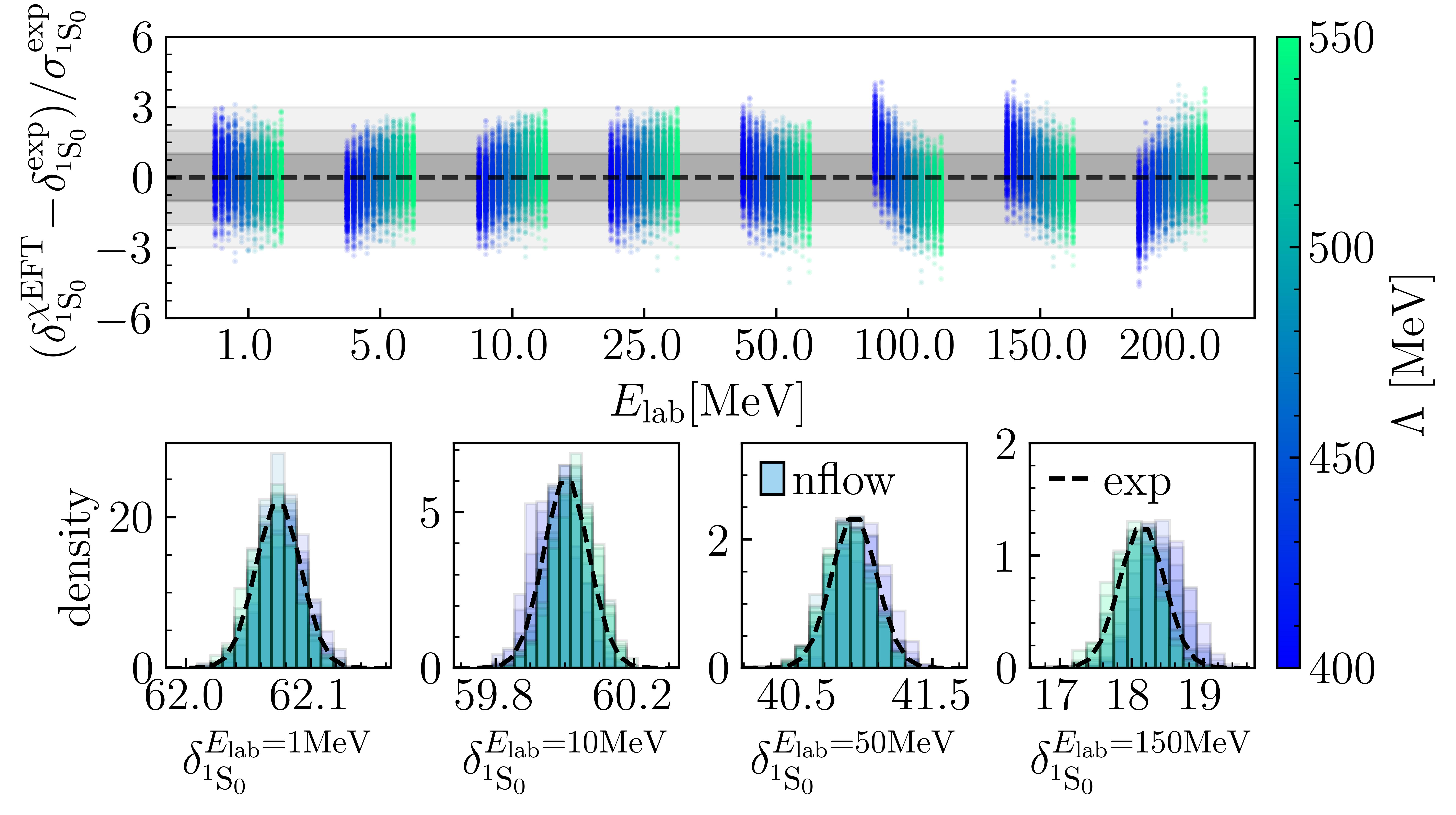}
    \caption{
    Distribution of phase shifts $^1{\rm S}_{0}$ whose LECs are generated by the trained normalizing flow as shown in Fig.~\ref{fig:rholec}. 
    The deviation (top panel) between the calculated phase shifts and experimental mean phase shift values across a wide range of $\Lambda$, normalized by the experimental standard deviation, is evaluated across all laboratory energies used to train the normalizing flow.
    The distributions of phase shifts (histogram) together with the experimental results (red lines), are shown at four laboratory energies.
    The experimental average and standard deviation are from the partial wave analysis in Ref.~\cite{NavarroPerez:2014ihw}.
    }
    \label{fig:rhophase}
\end{figure}

To assess the quality of the inferred LEC distributions, we propagate the samples through the LS equation and compare the resulting $^{1}\mathrm{S}_{0}$ phase-shift distributions directly with experiment. In the top panel of Fig.~\ref{fig:rhophase}, we show the deviations of the calculated phase shifts from the experimental mean, normalized by the corresponding experimental standard deviation, across all laboratory energies and resolution scales considered in the fit. The calculated phase shifts reproduce the experimental central values across the full range of $E_{\rm lab}$ and $\Lambda$, while their normalized deviations rarely exceed three standard deviations from. Representative phase-shift distributions at four laboratory energies, shown in the lower panels, further demonstrate that the widths of the experimental distributions are accurately reproduced and follow standard Gaussian distributions. Thus, the inferred LEC distributions capture not only the central values of the scattering observables but also their experimental uncertainties, providing a nontrivial validation of the strongly correlated parameter distributions shown in Fig.\ \ref{fig:rholec}.

\begin{figure}[b]
    \centering
    \includegraphics[width=\linewidth]{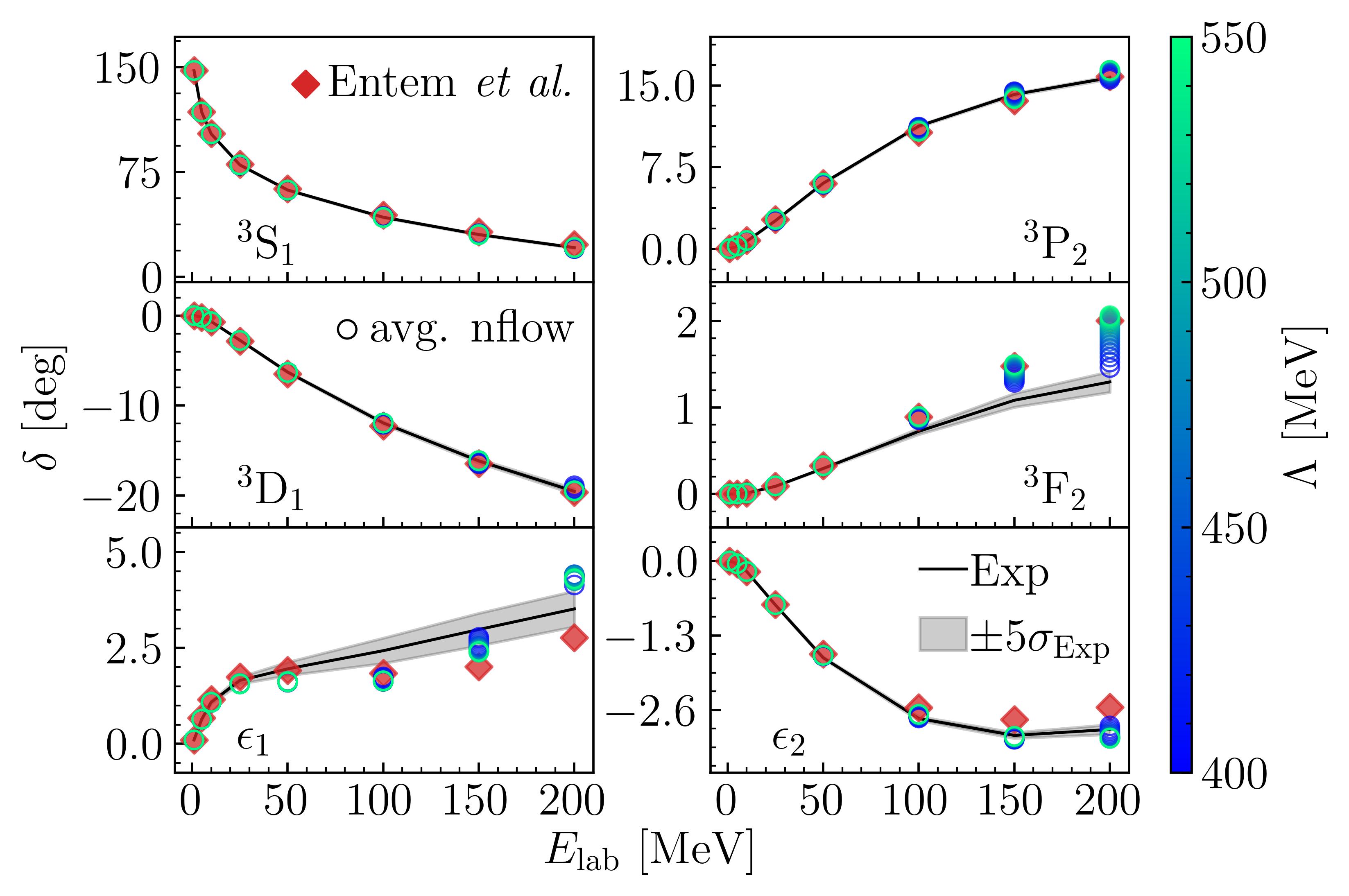}
    \caption{
    Mean phase shift values for coupled channels calculated from 500 LEC samples at each $\Lambda$ generated by the trained normalizing flow models (circles) are shown, with the experimental mean values (black lines) and error bands (black regions) from partial wave analysis \cite{NavarroPerez:2014ihw} as well as the non-linear regression fitting results (diamonds) for $\Lambda = 500 {\rm MeV}$ \cite{PhysRevC.91.014002} included for comparison.
    }
    \label{fig:phase}
\end{figure}

We also employ normalizing flows to construct LEC distributions for other phase-shift channels, covering total angular momenta from $J = 0$ to $3$ and a wide range of $\Lambda$ from 400 to 550 MeV. In Fig.~\ref{fig:phase}, we show the $\Lambda$-dependent mean phase shifts for the lowest coupled channels, including the mixing terms. In each case 500 LEC samples were generated at each $\Lambda$ from the well-trained normalizing flow models. In the $J=1$ coupled channel (left panel), solving the associated LS equation requires four blocks for the partial-wave channels of the potential, which at N3LO in $\chi$EFT involves 8 LECs. This leads to a joint distribution of the LECs across all the corresponding coupled partial-wave channels. For the $J=2$ coupled channel (right panel), only three LECs are present at N3LO in $\chi$EFT for the two partial waves, but the distribution of the three LECs is constrained by all the coupled-channel phase shifts, including $\delta_{^{3}{\rm P}_2}$, $\delta_{^{3}{\rm F}_2}$, and the mixing term $\delta_{\epsilon_2}$. 
The results closely track the experimental mean values, indicating that the normalizing flow can give statistically well-defined expectation values of LECs across a wide range of $\Lambda$.

\begin{figure}[t]
    \centering
    \includegraphics[width=\linewidth]{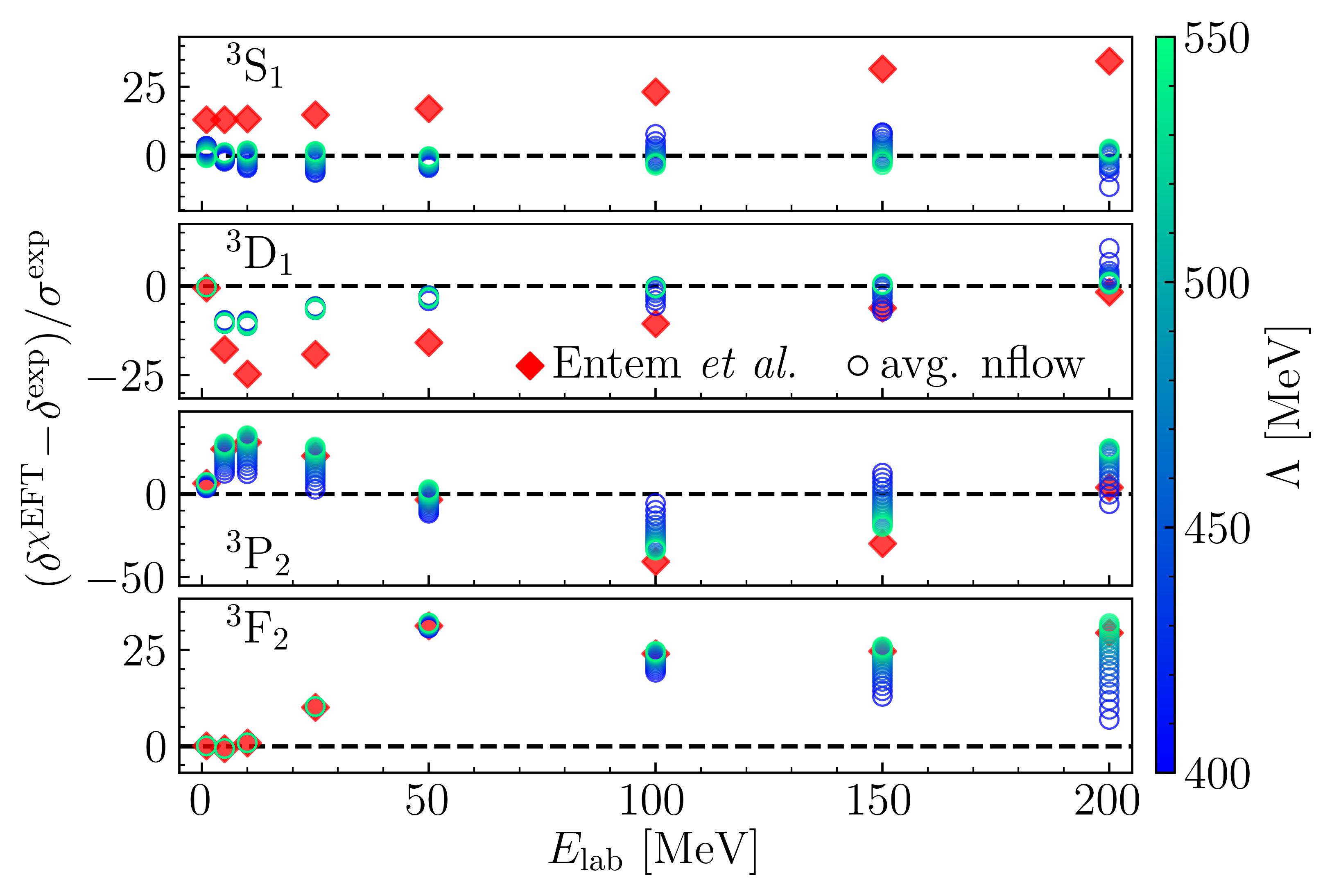}
    \caption{
    Difference between the theoretical mean and experimental phase shift values normalized by experimental uncertainties.
    The theoretical mean values at each $\Lambda$ are calculated based on 500 LEC samples from the trained normalizing flow.
    }
    \label{fig:phase_SPDF}
\end{figure}
In Fig.~\ref{fig:phase_SPDF}, we show the differences between the phase shifts obtained from the chiral interactions and the experimental values, normalized by the corresponding experimental uncertainties, for the coupled channels with $J=1,2$. Overall, the mean predictions are in good agreement with experiment, demonstrating that the normalizing flow is able to identify the expected value of LECs over a broad range of cutoff values $\Lambda$.
It is worth noting that there is no contact LEC specifically associated with the $^{3}{\rm F}_2$ partial wave at N3LO $\chi$EFT. Consequently, the $^{3}{\rm F}_2$ phase shifts exhibit a weaker dependence on the fitted LECs and a stronger dependence on the choice of resolution scale. This is reflected in the fact that the normalizing-flow predictions remain close to those obtained from the non-linear regression best-fit potential \cite{PhysRevC.91.014002} at $\Lambda=500{\rm MeV}$, even though the $^{3}{\rm F}_2$ channel is included in the loss function and therefore participates in the optimization in this work.
The agreement for these coupled channels is generally not as good as that obtained for the $^{1}{\rm S}_0$ channel. This may indicate limitations of the N3LO $\chi$EFT interaction itself with fewer fitting parameters in these channels, rather than deficiencies of the fitting procedure. Additional operator structures and higher-order dynamical contributions could increase the flexibility of the nuclear interaction and improve the description of the corresponding phase shifts.

{\it Summary and Outlooks:}
We have developed a generative framework for constructing statistical ensembles of chiral nuclear interactions across a continuous range of resolution scales. Rather than identifying a single optimal set of LECs at a fixed resolution scale, the normalizing flow learns their full correlated probability distribution conditioned on $\Lambda$, with the distribution constrained directly by experimental observables. Applied to neutron-proton scattering at N3LO, the framework efficiently identifies the regions of the multidimensional LEC space consistent with experimental phase-shift distributions over $\Lambda=400 - 550$\,MeV.

The resulting ensembles accurately reproduce both the central values and experimental uncertainties of the neutron-proton phase shifts across laboratory energies up to 200\,MeV. At the same time, the inferred LEC distributions reveal strong, non-Gaussian correlations that evolve continuously with resolution scale. These results demonstrate that normalizing flows can resolve complex structures in high-dimensional nuclear-interaction parameter spaces while retaining the full statistical information needed to propagate uncertainties to nuclear observables.

More broadly, this framework provides a pathway toward nuclear interactions constrained simultaneously by diverse experimental and observational data. Few- and many-body observables, including nuclear binding energies and neutron-skin thicknesses, can be incorporated directly into the inference of LEC distributions, while extensions to three-nucleon interactions and higher orders in $\chi$EFT would enable a more complete characterization of nuclear-force uncertainties \cite{epelbaumPrecisionNucleonNucleonPotential2015}. The resulting statistical ensembles can then be propagated to nuclear matter and astrophysical observables, providing a unified connection between microscopic nuclear interactions, laboratory experiments, and neutron-star observations.

\begin{acknowledgments}
The work of J.W.H. was supported by the National Science Foundation under Grant No.\ PHY-2514930.
The work of P.W. was supported by Texas A\&M Nuclear Solutions Institute, Cyclotron Institute.

\end{acknowledgments}

\vspace{\baselineskip}
{\it Data availability:} The data that support the findings of this article are openly available \cite{zenodo_21986908}.

\bibliography{cite}
\bibliographystyle{apsrev4-2}

\end{document}